\documentclass[12pt]{iopart}
\usepackage{iopams,setstack,epsfig}
\include{psfig}
\def\egal{\overset{\mathrm{def}}{=}}
\def\dfrac#1#2{\displaystyle\frac{#1}{#2}}
\def\binom#1#2{\left(#1\atop #2\right)}

\begin{document}

\title{Three-point correlations for quantum star graphs}
\author{Marie-Line Chabanol}
\address{Institut de Math\'ematiques de Bordeaux\\UMR 5251 CNRS-Universit\'e Bordeaux1\\France}
\ead{Marie-Line.Chabanol@math.u-bordeaux1.fr}

\begin{abstract}{We compute the three point correlation function for
    the eigenvalues of the Laplacian on 
    quantum star graphs in the limit where the number of edges tends
    to infinity. This extends a work by Berkolaiko and Keating, where
    they get the 2-point correlation function and show that it follows
    neither Poisson, nor random matrix statistics. It makes use of the
    trace formula and combinatorial analysis.}
\end{abstract}
\submitto{\JPA}

\section{Introduction}

The study of the Laplacian on a metric graph, a concept known as
{\it quantum graphs}, now serves as a toy model for quantum
chaos \cite{KS1, GS,BG}. Indeed, there exists an exact trace formula
relating eigenvalues and periodic orbits \cite{KS1,GS,KPS}. 
Moreover, depending on the
graphs, exact computations of these orbits may be possible, whereas
they are out of reach in most dynamical systems. It has thus been
shown \cite{KS1,KS2,GA} that spectral statistics of simple generic graphs
follow random matrix statistics when the size of the graph tends to
infinity, as expected for chaotic quantum systems. Star graphs (graphs
formed by a central vertex connected to $v$ other vertices by edges of
different lengths) play a
special role because of the high degeneracy of their periodic
orbits. As could be expected, this degeneracy breaks the random matrix
statistics  : this has been shown by  the computation of the two-point
correlation function (\cite{ThB,Ber}). Moreover it seems
reasonable to  
 expect that random matrix statistics would be
retrieved by glueing some star graphs together
(just by one edge). 
Hence star graphs can really be used as a toy model for
what degeneracy can induce on statistics, and how degeneracy can be
broken. Their simplicity makes exact results easier to obtain : the
trace formula for star graphs has been shown to converge under quite reasonable
assumptions (\cite{Winn}). Moreover, star graphs may also be considered
as a discrete version of Seba billiards (\cite{BK}) : indeed the 
eigenvalues of quantum star graphs and the energy levels of Seba billiards
are solutions of similar equations, so that their study can
also say something about continuous dynamics,  not only discrete one.
As a step further in the understanding of this model, we will here compute
 the three point correlation function of such graphs. Moreover, it is likely  
 to be a useful ingredient for the computation of the two point
 correlation function for glued star graphs. Glueing decreases degeneracy, so that one could expect 
another intermediate statistics
(double star graphs have been considered in \cite{KSG}, but their graphs are not the same as 
here : they have only two vertices). 

Spectral statistics for quantum graphs and the trace formula relating
them to periodic orbits will be recalled in the first part. The second
part will state the 2-point correlation function as obtained in
\cite{Ber}. The third part will present the computation of the 3-point
correlation function. Perspectives regarding glued graphs will be
given in the conclusion.

\section{Quantum graphs : eigenvalues and trace formula}

We will start by some vocabulary and notations.
Let $G=(E,V)$ be a graph with a metric structure : to each edge 
$(i,j)\in E\subset V\times V$
is assigned a length
$l_{ij}$, such that $l_{ij} = l_{ji}$; although the graph is supposed
to be non oriented,
that is $(i,j)\in E \Rightarrow (j,i)\in E$, and $l_{ij} = l_{ji}$, we
will consider the edges to be oriented : $(i,j)$ is different from
$(j,i)$,
 it really describes the edge going from $i$
to $j$. On each edge $(i,j)$, one can thus define a
coordinate $x$ such that $x=0$ corresponds to the vertex $i$, and
$x=l_{ij}$ corresponds to the vertex $j$. A {\em periodic orbit of length
$n$} is a set of $n$ edges $(p_1,\ldots,p_n)$ such that $p_i$ ends where
$p_{i+1}$ starts (as well as $p_n$ and
$p_1$). A periodic orbit is  called {\em primitive} if it is not the
repetition of a shorter periodic orbit. A primitive orbit repeated $r$
times is a non-primitive orbit with {\em repetition number $r$}. 
We will
denote by $v_j$ the {\em valence} of the vertex $j$, that is the number of
its neighbours.
On each edge $(i,j)$, one is looking for the spectrum of  the 
Laplacian. In other words, one wants to find $\lambda$ and $\psi_{ij}$
such that
$-\frac{\rmd^2\psi_{ij} }{\rmd x^2} = \lambda^2 \psi_{ij}(x)$. As one looks
for eigenfunctions defined on the whole graph, one imposes continuity
relations at each vertex, $\psi_{ij}(0) = \psi_{ik}(0)$. Moreover, the
function should have a unique value on a given point, regardless of
the sense of the edge it belongs to : hence one
wants $\psi_{ij}(x) = \psi_{ji}(l_{ij}-x)$. Finally, one imposes
 Neumann condition on each vertex $\sum_{j}
\frac{d\psi_{ij}}{dx}|_0 = 0$. It is then a simple exercise to check that the
eigenvalues $\lambda$ are the solutions of $\det(I-e^{-i\lambda L}
S) =0$, where $S$ and $L$ are $|E|\times|E|$ matrices : $L$ is
diagonal with the
length of each edge as diagonal element, and $S$ is defined by
$S_{(i,j),(j,k)} = -\delta_{i,k} + \frac{2}{v_j}$.

The trace formula as obtained in \cite{KS1} (a derivation specific to
star graphs is given in \cite{Winn})  states that if $d(\lambda) =
\sum_n \delta(\lambda - \lambda_n)$ is the spectral density, then
$d(\lambda) = \frac{L}{2\pi} + \frac{1}{\pi} \sum_{n} \sum_{p \in P_n}
\frac{l_p}{r_p} A_p \cos (\lambda l_p)$.
Here $L$ is the total length of all edges, $P_n$ is the set of all
periodic orbits of period $n$ up to cyclic reordering (that is
$p_0,p_1,p_2$ and $p_1,p_2,p_0$ are the same orbits), $l_p$ is the
length of the orbit, $r_p$ its repetition number, and $A_p =
\prod_{i=1}^n S_{p_{i},p_{i+1}}$.

\begin{figure}
\begin{center}
\scalebox{0.4}{\input{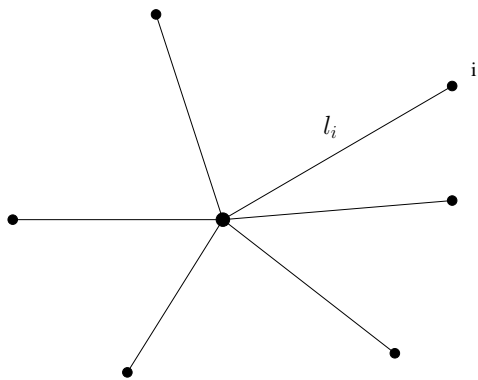}}
\caption{A star graph}
\label{stgraph}
\end{center}
\end{figure}

We will work on star graphs with $v+1$ vertices (see figure \ref{stgraph}):
these are  graphs with $V=\{0,..v\}$ and
$E=\{(0,i),(i,0), 1\leq i \leq v\}$ : $v$ vertices are all connected
to the center $0$. 
 The $S$-matrix elements are
$S_{(0,i),(i,0)} = 1$ (this corresponds to trivial scattering),
$S_{(i,0),(0,i)} = -1+\frac{2}{v}$ (backscattering) and
$S_{(i,0),(0,j)} = \frac{2}{v}$ (normal scattering).
The lengths will be taken so that they are incommensurate, and that
their distribution is peaked around 1 : for instance, they can be
chosen randomly, uniformly in
$[1-1/2v,1+1/2v]$, each length being independent from
the other ones. With such a distribution, an orbit of period $2k$ has a
length in $[2k-k/v,2k+k/v]$; such intervals for
differents $k$'s less than $v$ do not overlap.
The interesting limit will be the limit $v$ tends to infinity : in
this limit, orbits with the biggest contribution $A_p$ will be orbits
with a large number of backscatterings.

\section{2-point correlation function}

The 2-point correlation function is defined as $R_2(x) =\left(\frac{2\pi}{L}\right)^2 \langle d(\lambda) d(\lambda - \frac{2\pi
  x}{L})\rangle$. The brackets denote a mean value with respect  
to the $\lambda$'s, that is $\langle f\rangle
=\lim_{\Lambda \rightarrow \infty} \frac{1}{2\Lambda}
\int_{-\Lambda}^{\Lambda} {f(\lambda)} d\lambda$. 
Using the trace formula and performing the integral, one gets
\begin{eqnarray*}
R_2(x) =  1+\frac{2}{L^2} \sum_{p,p'}\frac{l_p l_{p'}}{r_p
r_{p'}} A_p A_{p'}  \delta_{l_p - l_{p'}}
\cos(\frac{2\pi x l_p}{L}),
\end{eqnarray*}
where the sum is over the pairs of periodic orbits $(p,p')$, up to cyclic
permutations, of  lengths $l_p$ and $l_{p'}$ and
repetition numbers $r_p$ and $r_{p'}$.

A combinatorial analysis of periodic orbits leads  (\cite{Ber})
to the formula \\
$R_2(x) = 1 + \int_{-\infty}^{+\infty} K(\tau) \exp(-2i\pi x \tau)
\rmd\tau$, where $K$ is given near $\tau =0$ by \\$K(\tau) = \exp(-4\tau) +
\sum_{j=2}^\infty \sum_{M=0}^\infty \frac{4^j}{j!} C_M
\tau^{M+j+1}$. The $C_M$ are defined by \\
$C_M = (-2)^M \sum_{k_1 + \ldots + k_j + n_1 + \ldots + n_j = M}
\frac{(K+j-1)! (N+j-1)!}{(M+j-1)!} \prod_{i=1}^j
\frac{\binom{n_i+k_i}{n_i}}{(n_i+1)!(k_i+1)!}$, with $K=\sum_{i=1}^j
k_i $ and $N= \sum_{i=1}^j n_i$.

It is found to be different from the Poisson statistics, since $K(\tau)$
clearly depends on $\tau$, and also different from  random matrix statistics,
since $K(0)=1$ and not 0.

\section{3-point correlation function}

The 3-point correlation function is defined in a similar way as
$ (\frac{2\pi}{L})^3 \langle d(\lambda)d(\lambda -
\frac{2\pi x}{L}) d(\lambda - \frac{2\pi y}{L})\rangle_{\lambda} 
$.
Using the trace formula and developing,  it is equal to
$
 R_2(x)+R_2(y) + R_2(x-y) - 2 + R_3(x,y)$, where \\
\begin{eqnarray*}
\fl R_3(x,y) = \left(\frac{2}{L}\right)^3 \lim_{\Lambda \rightarrow \infty}
\!\frac{1}{2\Lambda} \!\int_{-\Lambda}^\Lambda 
\!\sum_{p,p',p''}\!\frac{l_p l_{p'}l_{p''}}{r_p
r_{p'}r_{p''}} A_p A_{p'}A_{p''} \\\cos(\lambda l_p) \cos((\lambda \!-\!
\frac{2\pi x}{L})l_{p'}) \cos((\lambda \!-\! \frac{2\pi y}{L})l_{p''})
\rmd\lambda.
\end{eqnarray*}

 The sum is over  triplets of periodic orbits $(p,p',p'')$, up to cyclic
permutations, of  lengths $l_p, l_{p'}, l_{p''}$ and
repetition numbers $r_p,r_{p'},r_{p''}$. 
Performing the integral, one gets 

\begin{eqnarray*}
\fl R_3(x,y) =  \frac{2}{L^3} \sum_{p,p',p''}\frac{l_p l_{p'}l_{p''}}{r_p
r_{p'}r_{p''}} A_p A_{p'}A_{p''}\left[ \delta_{l_p + l_{p'} -l_{p''}}
\cos\left(\frac{2\pi}{L}(yl_{p''} -xl_{p'})\right)\right. \\\left.+ \delta_{l_p - l_{p'} +l_{p''}}
\cos\left(\frac{2\pi}{L}(yl_{p''} -xl_{p'})\right)+\delta_{l_p - l_{p'} -l_{p''}}
\cos\left(\frac{2\pi}{L}(yl_{p''} +xl_{p'})\right)\right]
\end{eqnarray*}

Let us have a look at the first term enforcing $l_{p''} = l_p +
l_{p'}$. The other ones can obviously be treated in a similar way.
We thus have to deal with 
$$R^1_3 =  \frac{2}{L^3} \sum_{{p,p',p''\atop l_{p''} = l_{p'}+l_p}}\frac{l_p l_{p'}(l_p+l_{p'})}{r_p
r_{p'}r_{p''}} A_p A_{p'}A_{p''}\cos\left(\frac{2\pi}{L}(yl_p
+(y-x)l_{p'})\right)$$

Now, since the edge lengths are incommensurate,  the
condition $l_{p''} = l_{p'}+l_p$ implies that the orbit $p''$ is
formed of the union of the edges of orbit $p$ and orbit $p'$, which we
will denote by $(p'')= (p)\cup(p')$.
Moreover, since the edge lengths are sharply peaked around 1, the length of a
periodic orbit of period $2k$ is nearly $2k$ and the total length $L$
is nearly $2v$.

Thus one gets  approximately
$$R^1_3 \simeq \frac{2}{v^3} \sum_{k,k'} kk'(k+k') 
\cos\left(\frac{4\pi}{L}(ky
+k'(y-x))\right)\sum_{{p \mathrm{\, period\, } 2k\atop p' \mathrm{\, period\, }
2k'}\atop (p'') = (p)\cup(p')} \frac{ A_p A_{p'}A_{p''}}{r_p
r_{p'}r_{p''}}$$

 Following
 \cite{Ber}, we will now take as a parameter the number $j$ of
distinct edges visited by the orbit $p''$.

\subsection{The $j=1$ case.}

The $j=1$ case is a bit special : the orbits $p,p'$
and $p''$ are here all formed of one and the same edge, for which
there are $\binom{v}{1}=v$ choices; the repetition
number of such an orbit of period $2k$ is then $r_p = k$, and the
number of backscatterings is also $k$, thus the
contribution $A_p$  is $(-1+\frac{2}{v})^k$.
Hence, the $j=1$ term is 
$$R^{1,1}_3 = \frac{2}{v^3}v\sum_{k,k'}  \cos\left(\frac{2\pi}{v}(ky
+k'(y-x))\right)\left(-1+\frac{2}{v}\right)^{2k+2k'}.$$

In the $v\rightarrow \infty$
limit,  putting $\tau = \frac{k}{v}$, the sum becomes an integral :
$$R^{1,1}_3 \simeq 2 \int \!\!\!\int_{{\mathbb R^+}^2} \rmd\tau \rmd\tau' \rme^{-4\tau} \rme^{-4\tau '}
\cos (2\pi (y \tau + (y-x)\tau ')).$$

\subsection{The $j=2$ case.}

This case is still a bit special, because while $p''$ passes through 2
distinct edges,  $p$ and $p'$ may still be restricted to one edge only.
The orbit $p''$ of period $2k''$ is here formed of 2 different edges, 
denoted by ``${\bf a}$'' and ``${\bf b}$'', for which there are
$\binom{v}{2}$ choices. It can be described as a succession of $m''$ packets
of ${\bf a}$ and ${\bf b}$. 
The number of scatterings and backscatterings of such
an orbit is respectively $2m''$ (each change of packet contributes)
and $k''-2m''$.   Its decomposition in $p\cup p'$ will obviously also
depend on the numbers $n''_a$ of edges ${\bf a}$ and $n''_b = k''-n''_a$ of
edges ${\bf b}$ (we count
the edges only when they depart from the root). 
One then needs to know how many different orbits
$p''$ there are with given $k'', m'', n''_a, n''_b$, since they will all contribute
the same. This number, as explained in \cite{Ber}, can be computed as
follows : you first divide your ${\bf a}$'s into $m''$ packets (order counts),
then your ${\bf b}$'s into $m''$ packets. Reminding that the number of
partitions of an integer $N$ into $K$ parts is $\binom{N-1}{K-1}$, the
number of ways to do this is $\binom{n''_a
-1}{m''-1}\binom{n''_b-1}{m''-1}$. The orbits should be counted up to
cyclic permutation, and the weight of each orbit has to be divided by
its repetition number $r$. But one orbit with a given $r$ corresponds
to $m''/r$ such decompositions. For example, the orbit ${\bf aabab}$ has
$m=2$, $r=1$, and actually we get it twice, since it is the same up to
cyclic permutation  as 
${\bf abaab}$. ${\bf abab}$ has $m=2$ and $r=2$, and it is obtained only once. 
Hence each decomposition gets a $1/m''$ factor, as well as a
contribution
 $A_p'' = (-1+\frac{2}{v})
^{k''-2m''} (\frac{2}{v})^{2m''}$.

Each such orbit has then to be decomposed into two orbits $p$ and $p'$,
composed of respectively $n_a$ ${\bf a}$'s and $n'_a = n''_a-n_a$
${\bf a}$'s (and
$n_b$ and $n'_b$ ${\bf b}$'s),
forming respectively $m$ and $m'$ packets.
The period of $p$ is $2(n_a +n_b)$.
One has to pay attention to the fact that
$n_a$ can be $0$, in which case $m=1$ but the number of scatterings is
then $0$ and not $2$ (this does not happen for $j>1$ in the case of the 2-point correlation function
computed in \cite{Ber} since 
there are then only two orbits $p$ and $p'$ visiting the same
edges). This term has to be computed separately.

Putting all that together, 
the term $\tilde R^{1,2}_3$ where $p$, $p'$ and $p''$  are all composed of
two different edges is :

\begin{eqnarray*}
\fl \tilde R^{1,2}_3 =
 \frac{2}{v^3}\! \binom{v}{2}\!\sum_{n''_a, n''_b}
 \sum_{{n_a=1\atop n'_a = n''_a -
n_a}}^{n''_a-1}
\sum_{{n_b=1\atop n'_b = n''_b-n_b}}^{n''_b-1}
\sum_{m=1}^{\min(n_a,n_b)}\sum_{m'=1}^{\min(n'_a,n'_b)}\sum_{m''=1}^{\min(n''_a,n''_b)}
(n''_a+n''_b) (n'_a + n'_b) (n_a + n_b)\\
\cos(\frac{2\pi}{v}((n_a+n_b)y + (y-x)(n'_a + n'_b)))  \left(-1+\frac{2}{v}\right)^{2n''_a+2n''_b-2(m+m'+m'')}\\
\frac{\binom{n''_a
-1}{m''-1}\binom{n''_b-1}{m''-1}}{m''} \frac{\binom{n'_a
-1}{m'-1}\binom{n'_b-1}{m'-1}}{m'}\frac{\binom{n_a
-1}{m-1}\binom{n_b-1}{m-1}}{m} 
\left(\frac{2}{v}\right)^{2(m+m'+m'')}.
\end{eqnarray*}

\noindent One can now perform the $v \rightarrow \infty$ limit :
denoting $q^*_i = \frac{n^*_i}{v}$ (the $*$ is either void, $'$ or
$''$), 
sums over $n$'s turn into
integrals over $q$'s , powers of $(1-\frac{2}{v})$
 turn into exponentials of $q$'s, terms such as
$\frac{(n-1)!}{(n-m)!}$  turn into $(qv)^{m-1}$. 
Hence, writing $q' = q''-q$ : 

\begin{eqnarray*}
\fl \tilde R_3^{1,2} \simeq
 \int_{({\mathbb R^+})^4} \rmd q_a \rmd q_b
\rmd q''_a \rmd q''_b  (q_a
+ q_b)(q''_a+q''_b)(q'_a+q'_b) \cos (2\pi ((q_a+q_b)y +
(y-x)(q'_a + q'_b))) \\  \exp(-4(q''_a + q''_b)) 4^3 \sum_{m''\geq
1} 
\frac{(4q''_aq''_b)^{m''-1}}{(m''-1)! m'' !} \sum_{m'\geq
1} 
\frac{(4q'_aq'_b)^{m'-1}}{(m'-1)! m' !} \sum_{m\geq
1} 
\frac{(4q_aq_b)^{m-1}}{(m-1)! m !}
\end{eqnarray*}
\noindent where the integral is over $\{q''_{a,b}\geq 0$ and $0\leq q_{a,b}\leq q''_{a,b}\}$.
Using the modified Bessel function $I_1(z) = (z/2) \sum_{k\in
\mathbb{N}} \frac{(z^2/4)^k}{k!(k+1)!}$, and denoting $\mathfrak{I}(x)
= I_1(4\sqrt x)/\sqrt x$, one gets :

\begin{eqnarray*}
\fl \tilde R_3^{1,2} \simeq 
%
%
8 \int\!\!\!\int_{({\mathbb R^+})^2} \rmd\tau \rmd\tau'  \rme^{-4\tau} \rme^{-4\tau '}
\cos (2\pi (y \tau + (y-x)\tau') \tau \tau' (\tau + \tau') \\  \int_0^\tau
\rmd q \int_0^{\tau'} \rmd q' \mathfrak{I}((q+q')(\tau+\tau' -q-q')) \mathfrak{I}(q(\tau-q)) \mathfrak{I}(q'(\tau' - q')).
\end{eqnarray*}

Let us now look at the case where one of the orbits $(p,p')$ is composed
of only one edge; for example,  the contribution of the term $n_a =
0$ (and thus $n'_a = n''_a$) is :
\begin{eqnarray*}
 \fl
 \frac{2}{v^3} \binom{v}{2}\sum_{n''_a, n''_b}
\sum_{m''=1}^{\min(n''_a,n''_b)} \!
\sum_{{n_b=1\atop n'_b = n''_b-n_b}}^{n''_b-1}\!
\sum_{m'=1}^{\min(n''_a,n'_b)}
(n''_a+n''_b) (n''_a + n'_b) \frac{\binom{n''_a
-1}{m''-1}\binom{n''_b-1}{m''-1}}{m''} \frac{\binom{n''_a
-1}{m'-1}\binom{n'_b-1}{m'-1}}{m'}\\
\cos\left(\frac{2\pi}{v}(n_b y + (y-x)(n''_a + n'_b))\right) 
 \left(-1+\frac{2}{v}\right)^{2(n''_a+n''_b-m'-m'')}
\left(\frac{2}{v}\right)^{2(m'+m'')}
\end{eqnarray*}

Since there are two symbols ${\bf a}$ and ${\bf b}$, and since there are two
orbits $p$ and $p'$ that can be degenerate, the total contribution
$\bar{R}_3^{1,2}$ of the ``one degenerate orbit'' case
is, when $v$ tends to infinity  :
\begin{eqnarray*}
\fl \bar{R}_3^{1,2} \simeq  8 \!\int \!\!\!\int_{({\mathbb R^+})^2} \rmd\tau \rmd\tau' \rme^{-4\tau} \rme^{-4\tau '}\!
\cos (2\pi (y \tau + (y-x)\tau '))  (\tau+\tau')\\
\left[\tau'
\int_0^{\tau'}\!\! \rmd q \mathfrak{I}(q(\tau+\tau'-q))
\mathfrak{I}(q(\tau'-q))
+\tau
\int_0^{\tau}\!\! \rmd q \mathfrak{I}(q(\tau+\tau'-q))\mathfrak{I}(q(\tau-q))\right]
\end{eqnarray*}

When both orbits $p$ and $p'$ consist of one edge (this problem is
specific to the $j=2$ case),  the contribution $\hat R^{1,2}_3 $ is :

\begin{eqnarray*}
\fl \hat R^{1,2}_3 = \frac{2}{v^3} \binom{v}{2}\sum_{n''_a, n''_b}
\sum_{m''=1}^{\min(n''_a,n''_b)} 
n''_a n''_b (n''_a+n''_b) 
\cos(\frac{2\pi}{v}(n''_b y + n''_a(y-x))) 
\frac{\binom{n''_a
-1}{m''-1}\binom{n''_b-1}{m''-1}}{m''} \frac{1}{n''_a n''_b} \\
\left(-1+\frac{2}{v}\right)^{2n''_a+2n''_b-2m''}
\left(\frac{2}{v}\right)^{2m''}\\
\simeq  \int \!\!\!\int_{({\mathbb R^+})^2} \rmd\tau \rmd\tau'  \rme^{-4\tau} \rme^{-4\tau '}
\cos (2\pi (y \tau + (y-x)\tau ') (\tau+\tau') \mathfrak{I}(\tau \tau').
\end{eqnarray*} 

All in all, this gives 
\begin{eqnarray*}
\fl R^{1,2}_3 = \tilde{R}^{1,2}_3 + \bar{R}^{1,2}_3 + \hat{R}^{1,2}_3
= \int \!\!\!\int_{({\mathbb R^+})^2} \rmd\tau \rmd\tau'  \rme^{-4\tau} \rme^{-4\tau '}
\cos (2\pi (y \tau + (y-x)\tau ') (\tau+\tau')\\
\left[ \mathfrak{I}(\tau \tau') + 8\tau \tau' \int_0^\tau
\rmd q \int_0^{\tau'} \rmd q' \mathfrak{I}((q+q')(\tau+\tau' -q-q'))
\mathfrak{I}(q(\tau-q)) \mathfrak{I}(q'(\tau' - q')) \right.\\
\left. + 8\tau'
\int_0^{\tau'}\!\! \rmd q \mathfrak{I}(q(\tau+\tau'-q))
\mathfrak{I}(q(\tau'-q))
+8\tau
\int_0^{\tau}\!\! \rmd q
\mathfrak{I}(q(\tau+\tau'-q))\mathfrak{I}(q(\tau-q)) \right] 
\end{eqnarray*}

\subsection{The $j>2$ case.}

The orbit $p''$ is now formed of $j$ edges, denoted by
$(\bf{1},\bf{2}, \ldots \bf{j})$ for which there are
$\binom{v}{j}$ choices. We will denote by $n''_i$ the number of edges
$\bf{i}$ in the orbit $p''$,  by $m''_i$ the number of groups of
adjacent ${\bf i}$, and by $\bf{n''}$ and ${\bf m''}$ the
corresponding vectors of ${\mathbb{Z}}^j$. For example, 
the orbit $p'' = {\bf 11212332}$ has
$j=3$, ${\bf n''} = (3,3,2)$, and ${\bf m''} = (2,3,1)$. 
The period of the orbit is $ 2\sum_{i=1}^j n''_i=2N''$, 
the number of scatterings is  the number of groups $M'' =
\sum_{i=1}^j m''_j$ and the number of backscatterings is
$\sum_{i=1}^j (n''_j-m''_j) = N''-M''$. We will denote by $Q_{{\bf
    n''}}^{{\bf m''}}$
the number of  orbits with given $j$, $\bf{n''}$ and ${\bf m''}$,
each weighted by $1/r_{p''}$.

Such an orbit has to be decomposed into two orbits $p$ and $p'$,
consisting respectively of $n_i$ and $n'_i= n''_i-n_i$ edges ${\bf i}$, and of
$m_i$ and $m'_i$ groups of ${\bf i}$. Some $n_i$ or $n'_i$ can be
zero, but not all of them, and 
as in the $j=2$ case, we will have to consider separately the case
where all the $n_i$ are zero but one.

To lighten the formulas, we will denote by $\prod^*  T^*
=   T  T' T''$ (where
$T$ can be any quantity we have defined for $p,p'$ and $p''$);
$\vec{1}\overset{\mathrm{def}}{=}(1,1,\ldots,1)$;
if $u$ and $v$ are vectors, $U \egal \sum_{i=1}^j u_i$, $\,u^v \egal 
\prod_{i=1}^j u_i^{v_i}$,  $\,u! \egal \prod_{i=1}^j (u_i)!$
 and  $u\leq
v$ means $u_i\leq v_i$ for all $i$.

In the general case where all orbits consist of at least two different
edges, we have :

\begin{eqnarray*}
\fl \tilde{R}_3^{1,j} = \frac{2}{v^3}\binom{v}{j}\sideset{}{'}{ \sum} NN''(N''-N)
\cos\left(\frac{2\pi}{v}(N''y -x(N''-N))\right) 
\\
\prod^* Q_{\bf n^*}^{\bf m^*}
\prod^*\left(-1+\frac{2}{v}\right)^{k^* -M^*}\left(\frac{2}{v}\right)^{M^*}
\end{eqnarray*}  

Here $\sideset{}{'}{\sum}$ denotes a sum over the  vectors
${\bf n^*}$ and ${\bf m^*}$ satisfying  :
$\cases{
\vec{0}\leq  {\bf n} \leq {\bf n''}\\
{\bf n'} = {\bf n'' - n}\\
\vec{1}\leq {\bf m^*}\leq {\bf n^*}}$

\noindent (Rigorously, we should avoid the case where  ${\bf n}$  or 
${\bf n'}$ is $\bf{0}$, but thanks to the $\frac{1}{v^3}$ term, 
its contribution  will disappear when the sums turn into integrals in the
$v\rightarrow \infty$ limit).
 
All we need now is to determine the numbers $Q_{{\bf n^*}}^{{\bf m^*}}$. The
computation is done in \cite{Ber}, let's just present the ideas
behind. We will count the sequences of $(\bf{1},\bf{2}, \ldots
\bf{j})$ such that there are $n_i \,\, \bf{i}$'s and $m_i$ groups of
$\bf{i}$, starting by a group of $\bf{1}$, and not ending by a group
of $\bf{1}$. Due to cyclic permutations, this is not exactly the same
as counting periodic orbits, but nearly : for example, the orbit ${\bf
11212332}$ corresponds to the $n_1/r=2$ sequences ${\bf 11212332}$ and
${\bf 12332112}$, whereas the orbit ${\bf 23112311}$ corresponds to the
only $n_1/r=1$ sequence ${\bf 11231123}$.  $Q$
is then exactly  the number of such sequences divided by $n_1$. 
To compute this number, one counts
the number of ways to put the $n_i$ ${\bf i}$'s in $m_i$ packets, and
then to arrange such packets, starting by {\bf 1}, keeping the order
of the groups of a given symbol, not ending with {\bf 1}, and in such
a way that two groups of the same symbol are not neighbours. 
The first step gives a factor
$\prod_{i=1}^j \binom{n_i-1}{m_i-1}$. The second step is the most
tricky one; it can be evaluated using
an exclusion/inclusion principle. All in all, this gives  

\begin{eqnarray*}
Q_{{\bf n}}^{{\bf m}} = \prod_{i=1}^j \binom{n_i-1}{m_i-1}
(-1)^M 
\sum_{\bf \vec{1}\leq t\leq m}
\frac{(-1)^T}{T}\binom{T}{t_1,\ldots, t_j}\prod_{i=1}^j
\binom{m_i-1}{t_i-1}
\end{eqnarray*}

Now all one has to do is to perform the $v\rightarrow \infty$
limit. Using $\sum_{m=1}^\infty \sum_{t
=1}^{m} = \sum_{t=1}^{\infty} \sum_{m=t}^{\infty}$ and
$\sum_{m=t}^{\infty} \frac{x^{m-1}}{(m-t)!} = x^{t-1}\exp x$
and introducing $\tau = \sum_{i=1}^j q_j$, one gets 

\begin{eqnarray*}
\fl \tilde{R}_3^{1,j} \simeq 
\frac{2}{j!}
\sum_{{\bf t,t',t''}}   \int_{{\bf q''\geq} \vec{0}} \prod_{i=1}^j
\rmd q''_j
\int_{\bf 0\leq q \leq q''} \prod_{i=1}^j \rmd q_j  \tau
\tau''(\tau''-\tau)
\cos(2\pi(\tau x + \tau''(y-x)))\\
 \sideset{}{*}\prod {\bf (q^*)^{t^*-1}}
 \frac{(T-1)!(T'-1)!(T''-1)!}{\sideset{}{*}\prod {\bf t^*!(t^*-1)!}}(-2)^{T+T'+T''}
\end{eqnarray*}  

Using the identity $\int_{q_i\geq 0, \sum_{i=1}^j q_i = \tau}
\rmd q_1 \!\cdots \rmd q_{j-1} \prod_{i=1}^j q_i ^{m_i} = \dfrac{\prod_{i=1}^j
  m_i!}{(M + j-1)!} \tau^{M +j-1}$ and developing ${\bf (q'')^{t''-1}} =
{\bf (q+q')^{t''-1}}$, this becomes

\begin{eqnarray*}
\fl \tilde{R}_3^{1,j} \simeq
\frac{2}{j!} \int\!\!\int_{\mathbb{R^+}^2} \rmd\tau \rmd\tau'(\tau+\tau')
 \cos(2\pi(y\tau + (y-x)\tau')) \sum_{{\bf t}\geq \vec{1}} \sum_{{\bf
     t'} \geq \vec{1}}
\sum_{{\bf t''} \geq \vec{1}}\sum_{\vec{0} \leq {\bf s} < {\bf t''}}
\\
\prod_{i=1}^j\frac{ \binom{s_i+t_i-1}{s_i}
\binom{t''_i+t'_i-s_i-2}{t'_i-1}}
{ (t_i)!(t'_i)!(t''_i)!}
\frac{(T-1)!(T'-1)!(T''-1)!}{(S+T-1)!(T''+T'-S-j-1)!}
\\ (-2)^{T+T'+T''} \tau^{S+T}(\tau')^{T'+T''-S-j}
\end{eqnarray*}

The case where  $p$ consists of only one
edge gives a factor
\begin{eqnarray*}
\fl \frac{2j}{v^3}\binom{v}{j}{\sum^2} N''(N''-N)
\cos\left(\frac{2\pi}{v}(N''y -x(N''-N))\right) \\Q_{\bf n''}^{\bf m''}Q_{\bf
  n'}^{\bf m'}
\left(-1+\frac{2}{v}\right)^{2N'' -M'-M''}\left(\frac{2}{v}\right)^{M''+M'}
\end{eqnarray*}  

\noindent where ${ \sum^2}$ denotes a sum over the vectors
${\bf n''}$, ${\bf m''}$,  ${\bf m'}$ and over the integer $n_1$ satisfying :
\\$
\cases{
1\leq  n_1 \leq n''_1\\
\vec{1}\leq {\bf m''}\leq {\bf n''}\\
\vec{1}\leq  {\bf m'} \leq {\bf n''} \\
1\leq  m'_1 \leq n''_1-n_1
}
$\\
Hence, in the limit $v\rightarrow \infty$, the contribution of $p$ or
$p'$ consisting of one edge only is :
\begin{eqnarray*}
\fl \bar{R}_3^{1,j} \simeq 
   \frac{2}{(j-1)!} \int\!\!\int_{\mathbb{R^+}^2} \rmd\tau \rmd\tau'
(\tau+\tau') \cos(2\pi(y\tau + (y-x)\tau')) 
\\  \sum_{t'_i,t''_i} \sum_{s=0}^{t''_1-1} 
\frac{\binom{s+t'_i-1}{s} \prod_{i=2}^j
  \binom{t''_i+t'_i-2}{t'_i-1}}{(t''_i-1-s)! \prod_{i=1}^j (t'_i)!
  \prod_{i=1}^j (t''_i)! } \frac{(T'-1)!(T''-1)!}{(T''+T'-t''_1+s-j)!} (-2)^{T'+T''}
\\ \left[(\tau')^{T''+T'-t''_1+s-j+1} \tau^{t''_1-1-s} \exp(-2\tau)+
 \tau^{T''+T'-t''_1+s-j+1} (\tau')^{t''_1-1-s}\exp(-2\tau')\right]
\end{eqnarray*}  

Finally, the contribution for $j>2$ is  $R_3^{1,j} = \tilde{R}_3^{1,j} + \bar{R}_3^{1,j}$.
\subsection{The 3-point correlation function}
Putting everything back together,
the 3-point correlation function can  be written 
$R_2(x) + R_2(y) + R_2(x-y) - 2 +\int \!\!\!\int_{{\mathbb R^+}^2} \rmd\tau \rmd\tau' (\cos (2\pi(y\tau  + (y-x)\tau')+
\cos (2\pi(y\tau' - x(\tau+\tau')) + \cos (2\pi(y\tau +x\tau'))  F(\tau,\tau')
 $, where $F=F_1+F_2+F_3+F_4$ is given by
$F_1(\tau,\tau') = 2 \rme^{-4\tau} \rme^{-4\tau '}$, 
\begin{eqnarray*}
\fl F_2(\tau,\tau')=
\rme^{-4\tau} \rme^{-4\tau'}(\tau+\tau')\left[\mathfrak{I}(\tau \tau') 
+ 8
\tau'
\int_0^{\tau'}\!\! \rmd q \mathfrak{I}(q(\tau+\tau'-q))
\mathfrak{I}(q(\tau'-q))\right.\\ 
 +8\tau
\int_0^{\tau}\!\! \rmd q
\mathfrak{I}(q(\tau+\tau'-q))\mathfrak{I}(q(\tau-q))\\
+\left. 8\tau \tau' \int_0^\tau
\rmd q \int_0^{\tau'} \rmd q' \mathfrak{I}((q+q')(\tau+\tau' -q-q'))
\mathfrak{I}(q(\tau-q)) \mathfrak{I}(q'(\tau' - q'))\right] 
\end{eqnarray*}
\begin{eqnarray*}
 \fl F_3(\tau,\tau') = \sum_{j\geq 3}\frac{2}{j!} (\tau+\tau')
  \sum_{{\bf t}\geq \vec{1}} \sum_{{\bf
     t'} \geq \vec{1}}
\sum_{{\bf t''} \geq \vec{1}}\sum_{\vec{0} \leq {\bf s} < {\bf t''}}
\prod_{i=1}^j\frac{ \binom{s_i+t_i-1}{s_i}
\binom{t''_i+t'_i-s_i-2}{t'_i-1}}
{ (t_i)!(t'_i)!(t''_i)!}\\
\frac{(T-1)!(T'-1)!(T''-1)!}{(S+T-1)!(T''+T'-S-j-1)!}
 (-2)^{T+T'+T''} \tau^{S+T}(\tau')^{T'+T''-S-j}
\end{eqnarray*}
\begin{eqnarray*}
\fl F_4(\tau,\tau') = \sum_{j\geq 3}    \frac{2}{(j-1)!}
(\tau+\tau')  \sum_{t'_i,t''_i} \sum_{s=0}^{t''_1-1} 
\frac{\binom{s+t'_i-1}{s} \prod_{i=2}^j
  \binom{t''_i+t'_i-2}{t'_i-1}}{(t''_i-1-s)! \prod_{i=1}^j (t'_i)!
  \prod_{i=1}^j (t''_i)! } \\
\frac{(T'-1)!(T''-1)!}{(T''+T'-t''_1+s-j)!} (-2)^{T'+T''}
 \left[(\tau')^{T''+T'-t''_1+s-j+1} \tau^{t''_1-1-s} \exp(-2\tau)\right.\\\left.+
 \tau^{T''+T'-t''_1+s-j+1} (\tau')^{t''_1-1-s}\exp(-2\tau')\right]
\end{eqnarray*}

\subsection{Small $\tau, \tau'$ expansion.}

\begin{figure}
\begin{center}
$\begin{array}{cc}
\epsfig{file=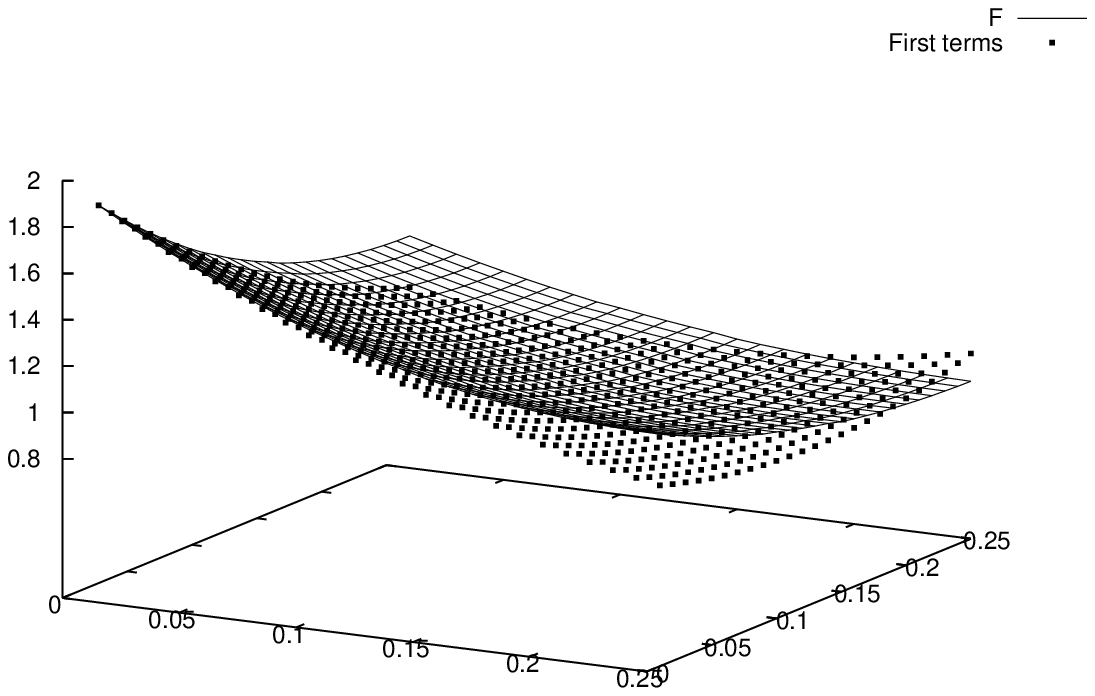,height=5.5cm}&
\epsfig{file=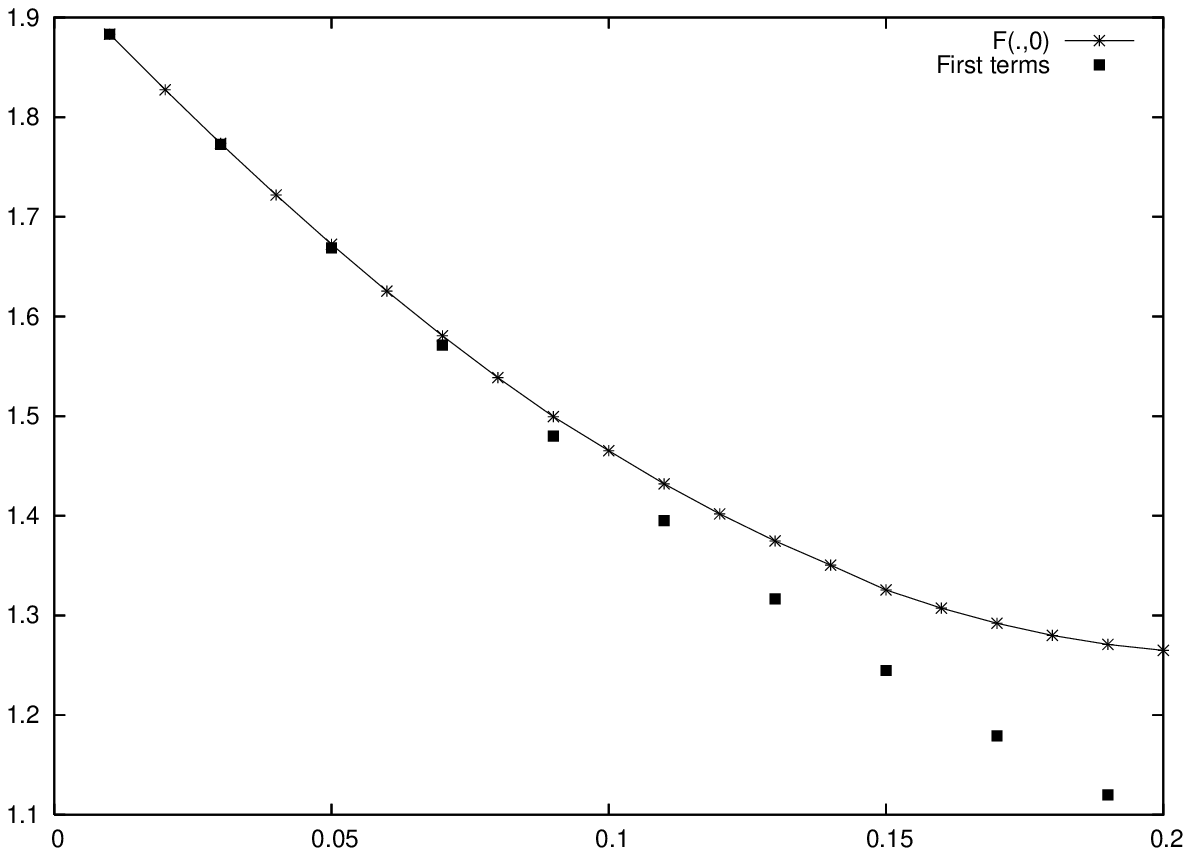,height=5cm}\\
(a)&(b)
\end{array}$
\caption{The function F (line) and its expansion up to order 2
  (squares); (b) corresponds to $\tau'=0$.}
\label{figF}
\end{center}
\end{figure}

 The first contribution for the ``general'' $j$-th term
where no orbit consists of only one edge
 is $(\tau \tau')^j(\tau+\tau')$, and the first contribution
when one orbit is degenerate is $(\tau')^j(\tau+\tau')$. Keeping
only the first terms,  and using the expansion of $\mathfrak{I}(x) = 2+4x +O(x^2)$ when $x$ is
small, one thus gets $F(\tau,\tau') =
2-6\tau-6\tau'+16\tau\tau' +8\tau^2+8(\tau')^2 +
o(\tau^2,\tau'^2,\tau\tau')$. Figure \ref{figF} presents a graph of
the function $F$, together with the first terms of its expansion.

\section{Conclusion and perspectives}

We have obtained here a formula for the 3-point correlation function
of star graphs, allowing in principle to get its small $\tau,\tau'$
expansion at any fixed order $s$ by keeping terms up to $j=s-1$.  
One should be able to get the first order of the expansion for the
$n$-point correlation function in the same way by computing the 
$j=1$ and $j=2$ contributions. Since the statistics of the eigenvalues
are characterized by all the $n$-point correlation functions, this is
a small step towards knowing a bit more of this intermediate
statistics for star graphs.

This specific 
result for the 3-point function could also be of some help in computing the form factor of 
 two star graphs $S_1$ and $S_2$ 
 glued together by  an edge
${\bf e}$
linking their centers. Indeed,
the form factor is a sum over pairs of orbits $(p,q)$ 
visiting the same
edges. Let us consider all the couples $(p,q)$ 
corresponding to a given set of edges. To get the main term in the small $\tau$ expansion, 
it would seem reasonable to group the edges corresponding to each graph together, and thus to consider first 
the couples $(p,q)$  such that $p$ and $q$ can be written
 $p=p_1Ep_2E$ and $q=q_1Eq_2E$, where $p_i$ and $q_i$ are orbits on $S_i$, 
and $E$ denotes any sequence ${\bf e ... e}$ of edges ${\bf e}$. The sum over such $p$ and $q$ should  
thus involve a sum over $(p_1,q_1)$ and a sum over $(p_2,q_2)$, and 
the product of the form factors of each star graph should appear here. 
The next term would correspond to decompositions $p=p_1Ep_2Ep'_1Ep'_2$ and $q = q_1Eq_2E$. 
 The sum over orbits $p,q$ would then involve
 a sum  on each star graph over orbits $p_i,p'_i,q_i$ such that
$(q_i)=(p_i)\cup (p'_i)$,  similar to the sum computed in the three-point correlation function.
   Of course, the exact
calculation would involve  all the $n$-point correlation
functions as well as combinatorial factors to insert the $E$'s, but
the first terms of the expansion should already give an insight of
what happens.


\section{Acknowledgments}

The author is grateful to the referees whose comments were of great help to improve the presentation of the 
paper. 

\section*{References}
\bibliographystyle{unsrt}
\bibliography{qgraph}

\begin{thebibliography}{10}

\bibitem{KS1}
T.~Kottos and U.~Smilansky.
\newblock Periodic orbit theory and spectral statistics for quantum graphs.
\newblock {\em Ann. Phys.}, 274:76--124, 1999.

\bibitem{GS}
S.~Gnutzmann and U.~Smilansky.
\newblock Quantum graphs : applications to quantum chaos and universal spectral
  statistics.
\newblock {\em Advances in Physics}, 55:527--625, 2006.

\bibitem{BG}
F.~Barra and P.~Gaspard.
\newblock On the level spacing distribution in quantum graphs.
\newblock {\em J. Stat. Phys.}, 101:283--319, 2000.

\bibitem{KPS}
V.~Kostrykin, J.~Potthoff, and R.~Schrader.
\newblock Heat kernels on metric graphs and a trace formula.
\newblock {\em arXiv:math-ph/0701009}, 2007.

\bibitem{KS2}
T.~Kottos and U.~Smilansky.
\newblock Quantum graphs : a simple model for chaotic scattering.
\newblock {\em J. Phys. A}, 36:3501--3524, 2003.

\bibitem{GA}
S.~Gnutzmann and A.~Altland.
\newblock Universal spectral statistics in quantum graphs.
\newblock {\em Phys. Rev. Lett.}, 93, 2004.

\bibitem{ThB}
G.~Berkolaiko.
\newblock {\em Quantum star graphs and related systems}.
\newblock PhD thesis, University of Bristol, 2000.

\bibitem{Ber}
G.~Berkolaiko and J.~P. Keating.
\newblock Two point spectral correlations for star graphs.
\newblock {\em J. Phys. A}, 32:7827--7841, 1999.

\bibitem{Winn}
B.~Winn.
\newblock On the trace formula for quantum star graphs.
\newblock In G.~Berkolaiko, R.~Carlson, S.~A. Fulling, and P.~Kuchment,
  editors, {\em Quantum graphs and their application}, volume 415, pages
  293--307. AMS, 2006.

\bibitem{BK}
G.~Berkolaiko, E.~B. Bogomolny, and J.~P. Keating.
\newblock Star graphs and {S}eba billiards.
\newblock {\em J. Phys. A}, 34:335--350, 2001.

\bibitem{KSG}
M.~Kopp, H.~Schanz, and T.~Geisel.
\newblock The spectral form factor of double star graphs.
\newblock Poster, 2005.

\end{thebibliography}
\end{document}